\documentclass[10pt]{article}
\ifx\pdfpagewidth\undefined
 \else %% just for pdf output
\fi
\usepackage{amsfonts, amsbsy, amssymb, amsmath} 
\begin{document}
%%===============================================
\title{Parameter Estimation from Censored Samples 
       using the Expectation-Maximization Algorithm}

\author{
Chanseok Park  \\
%% \thanks{Email: \texttt{cspark@ces.clemson.edu}}\\
Department of Mathematical Sciences\\
Clemson University \\
Clemson, SC 29634 \\
USA
\and
Seong Beom Lee \\
School of Mechanical \& Automotive Engineering\\
Inje University \\
Kimhae, Kyongnam\\
South Korea}

\date{April 2, 2003}
\maketitle
%============================
%---------------
\begin{abstract}
This paper deals with parameter estimation
when the data are randomly right censored.
The maximum likelihood estimates from censored samples are obtained 
by using the expectation-maximization (EM)
 and Monte Carlo EM (MCEM) algorithms.
We introduce the concept of the EM and MCEM algorithms and
develop parameter estimation methods
for a variety of distributions
such as normal, Laplace and Rayleigh distributions.
These proposed methods are illustrated with three examples.

%---------------
%%\begin{keyword}
\end{abstract}
%============================

\clearpage

%============================
\section{{Introduction}}
%============================
The analysis of lifetime or failure time data
has been of considerable interest 
in many branches of statistical applications such as 
electrical engineering, biological sciences, medicine, etc.
In reliability analysis, censoring is very common because of 
time and cost considerations on experiments. 
The data are said to be censored when, for observations, only a lower
or upper bound on lifetime is available.

The problem of parameter estimation from censored samples 
has been treated by several authors.
Gupta~\cite{Gupta:1952} has studied
the maximum likelihood estimate (MLE) and the best linear estimate
%%% the MLE and the best linear estimate 
for Type-I and Type-II censored samples 
from a normal distribution.
To obtain the MLE, some numerical optimization techniques need to be
employed.
Unless otherwise specified, ``MLE'' refers to the estimate obtained
by direct maximization of the likelihood function.
Maximizing the likelihood function directly using gradient methods such as
the Newton-Raphson method is very fast and efficient.
However, these methods are very sensitive to starting values.
Also, the larger the number of parameters, the higher the chance 
that the method will be unable to find the maximizer.
In addition, if the likelihood function is very flat near its maximum, 
then the method will stop before finding the maximizer.
Sultan~\cite{Sultan:1997} has given an approximation of the MLE
for a Type-II censored sample from a normal distribution.
Govindarajulu~\cite{Govindarajulu:1966} has derived
the best linear unbiased estimate (BLUE) for a symmetrically 
%% the BLUE for a symmetrically
Type-II censored sample from a Laplace distribution for $N$ up to 20.
Balakrishnan~\cite{Balakrishnan:1996} has given the 
BLUE for a Type-II censored sample from a Laplace distribution.
The BLUEs need the coefficients  $a_i$ and $b_i$, which 
were tabulated in \cite{Balakrishnan:1996}, but 
the table is provided only for sample size $N=20$.
The approximate MLE and the BLUE do not converge to the MLE.
The methods above are also restricted only to Type-I or Type-II
censored samples.
These deficiencies can be overcome by the proposed methods
based on the EM and MCEM algorithms.

Until the advent of powerful and accessible computing methods,
the experimenter was often confronted with a difficult choice: 
Either describe an accurate model of a phenomenon, which
would usually preclude the computation of explicit answers, or
choose a feasible approximate model 
which would avoid computational difficulties,
but may not be close to an exact model.
In an era of powerful computers, simulation-based estimation such as
the EM algorithm promises to be one of the mainstays of 
applied parametric modeling and data analysis in the years ahead.

We develop parameter estimation methods
via the EM and MCEM algorithms when the data are randomly right censored.
This random censoring is a generalization of Type-I censoring.
These MLEs are obtained using the EM and MCEM algorithms.
Since the calculation of the MLEs in Type-I and Type-II censoring
are nearly identical, the proposed methods can be used for 
Type-II censoring also.
We present the estimation methods when the data come from 
the normal, Laplace and Rayleigh distributions

In section~\ref{SEC:Model}, we introduce the likelihood construction
for censored data.
Section~\ref{SEC:EM} introduces the concept of the EM and MCEM algorithms. 
Section~\ref{SEC:Parameter} provides estimation procedure. 
Section~\ref{SEC:Examples} illustrates examples.

%============================
\section{{Likelihood construction for censored data}\label{SEC:Model}}
%============================
Suppose that 
we observe ${\mathbf{x}}=(x_1,\ldots,x_n)$
which are independent and identically distributed (iid)
and have a continuous distribution with
the probability density function (pdf) $f(x)$
and the cumulative distribution function (cdf) $F(x)$.
Data from experiments involving random censoring can be
conveniently represented by pairs $(w_i,\delta_i)$ 
with $w_i=\min(x_i,R_i)$:  
$$
\delta_i = \left\{ \begin{array}{r@{\qquad}l}
                 0 & x_i > R_i \\ 1 & x_i \le R_i
                 \end{array} \right.
\qquad{\mathrm{for}}\quad i=1,\ldots,n,
$$
where $\delta_i$ is a censoring indicator variable and 
$R_i$ is a censoring time of test unit $i$.
Denote the vector of unknown parameters 
by ${\boldsymbol{\theta}}=(\theta_1,\ldots,\theta_p)$.
Then ignoring an normalizing constant,
we have the complete-data likelihood
$$
L^c({\boldsymbol{\theta}} | {\mathbf{x}})
= \prod_{i=1}^{n} f(x_i).
$$
Denote the observed (uncensored) part of $x_1,\ldots,x_n$ by
${\mathbf{y}}=(y_1,\ldots,y_m)$ and the missing (censored) part by
${\mathbf{z}}=(z_{m+1},\ldots,z_{n})$ with $z_i>R_i$.
Integrating $L^c({\boldsymbol{\theta}} | {\mathbf{x}})$
with respect to ${\mathbf{z}}$,
we obtain the observed-data likelihood 
\begin{align*}
L({\boldsymbol{\theta}}|{\mathbf{y}})
&= \int L^c ({\boldsymbol{\theta}}|{\mathbf{y,z}}) d{\mathbf{z}} \\
&= \prod_{i=1}^{m} f(y_i) 
   \prod_{j=m+1}^{n} \int_{z_j>R_j} \!\!\!\!\!  f(z_j)\, d z_j   \\
&= \prod_{i=1}^{m} f(y_i) 
   \prod_{j=m+1}^{n} \big[1-F(R_j)\big]. 
\end{align*}
Using the $(w_i,\delta_i)$ notation, 
we have 
\begin{equation} \label{EQ:likelihood}
L({\boldsymbol{\theta}} | {\mathbf{w}},{\boldsymbol{\delta}})
= \prod_{i=1}^{n} [f(w_i)]^{\delta_i} [1-F(w_i)]^{1-\delta_i},
\end{equation}
where ${\mathbf{w}}=(w_1,\ldots,w_n)$ and 
      ${\boldsymbol{\delta}}=(\delta_1,\ldots,\delta_n)$.

For Type-II censoring, the data consist of the $r$th smallest
lifetimes $x_{(1)} \le x_{(2)} \le \cdots \le x_{(r)}$
out of a sample of size $n$.
Assuming that we observe ${\mathbf{x}}=(x_1,\ldots,x_n)$
which are iid 
and have a continuous distribution,
it follows that the joint pdf of $x_{(1)},\ldots,x_{(r)}$
(see \cite{David:1981}) is
$$
\frac{n!}{(n-r)!} \;
\prod_{i=1}^{r} f(x_{(i)})
\prod_{j=r+1}^{n}  [1-F(x_{(r)})].
$$
Ignoring  an normalizing constant, we can 
rewrite the above equation
 in the form of (\ref{EQ:likelihood}) 
by setting $R_i=x_{(r)}$.
Hence we can use (\ref{EQ:likelihood}) for the calculation
of the MLE in both Type-I (including random censoring)
 and Type-II censoring.

%============================
\section{{The EM and MCEM algorithms}\label{SEC:EM}}
%============================
% Schafer p.3
The EM algorithm
 is a general technique for finding maximum
likelihood estimates for parametric models when the data 
are not fully observed.
%%pg. 213 (Rober/Casella)
The EM algorithm was originally
introduced by Dempster {\em et al.}~\cite{Dempster/Laird/Rubin:1977}
 to overcome the difficulties in maximizing likelihoods.

% Schafer p.37
The key idea behind EM algorithm is to solve a difficult incomplete-data
problem by repeatedly solving tractable complete-data problems.
The E-step of each iteration only involves taking expectations over
complete-data conditional distributions and the M-step of each iteration
only requires complete-data maximum likelihood estimation, which
is often in simple closed form.
%% Schafer pg.51
For incomplete-data problems, the most attractive features 
of the EM algorithm relative to other optimization techniques are its
simplicity and its stability.
Rather than maximizing the potentially complicated likelihood
function of the incomplete data directly, 
we repeatedly maximize the log-likelihood function 
of the complete data given the incomplete data,
which is typically much 
easier and often equivalent to finding MLEs with complete data.
Moreover, successive iterations of the EM algorithm are guaranteed
never to decrease the likelihood function, which is not generally
true of gradient methods like Newton-Raphson.
Hence in the case of the unimodal and concave likelihood function,
the EM algorithm converges to the global maximizer from any starting value.
We can employ this methodology for parameter estimation 
from a censored sample 
since censored data models are special cases of missing data models.

The EM algorithm consists of two distinct steps:

%---------------------------------
\medskip 
\itemindent -0em \leftmargini 2em
\begin{itemize}  
\item \textsf{E-step:} \texttt{compute} \\
$Q({\boldsymbol{\theta}}|{\boldsymbol{\theta}}^{(s)})
   = \int\log L^c ({\boldsymbol{\theta}}|{\mathbf{y,z}})
  p({\mathbf{z}}|{\mathbf{y}},{\boldsymbol{\theta}}^{(s)})d{\mathbf{z}}$.
\medskip
\item \textsf{M-step:} \texttt{find} the ${\boldsymbol{\theta}}^{(s+1)}$ \\
which maximizes $Q({\boldsymbol{\theta}}|{\boldsymbol{\theta}}^{(s)})$ 
in ${\boldsymbol{\theta}}$.
\end{itemize}  
\medskip

A difficulty with the implementation of the EM algorithm is 
that each \textsf{E-Step} requires the integration of the 
expected log-likelihood  to obtain the 
$Q({\boldsymbol{\theta}}|{\boldsymbol{\theta}}^{(s)})$.
Because of the integration, maximizing 
$Q({\boldsymbol{\theta}}|{\boldsymbol{\theta}}^{(s)})$
can be difficult even when maximizing
$L^c({\boldsymbol{\theta}}| {\mathbf{y,z}})$
is trivial.
Wei and Tanner~\cite{Wei/Tanner:1990a,Wei/Tanner:1990b}
propose using the MCEM to overcome this difficulty
by simulating $z_{m+1},\ldots,z_{n}$ from
the conditional distribution 
$p({\mathbf{z}}|{\mathbf{y}},{\boldsymbol{\theta}}^{(s)})$
and then maximizing the approximate
expected log-likelihood
$$
\hat{Q}({\boldsymbol{\theta}}|{\boldsymbol{\theta}}^{(s)})
=\frac{1}{K}\sum_{k=1}^{K} 
 \log L^c ({\boldsymbol{\theta}}|{\mathbf{y}},{\mathbf{z}}^{(k)}),
$$
where ${\mathbf{z}}^{(k)}=(z_{m+1,k},\ldots,z_{n,k})$.

The books by 
Little and Rubin~\cite{Little/Rubin:2002}, Tanner~\cite{Tanner:1996},
and Schafer~\cite{Schafer:1997}
provide good overview of the EM literature.

%============================
\section{{Parameter estimation}\label{SEC:Parameter}}
%============================

%----------------------------
\subsection{The normal distribution}
%----------------------------
Let $Y_1,\ldots,Y_m$ and $Z_{m+1},\ldots,Z_n$ be 
iid normal random variables with 
$\boldsymbol{\theta}=(\mu,\sigma^2)$. 
Then the complete-data log-likelihood is 
\begin{align*}
&\log L^c({\boldsymbol{\theta}} | {\mathbf{y,z}})   \\
&= C - \frac{n}{2}\log\sigma^2 - \frac{1}{2\sigma^2}
 \big\{\sum_{i=1}^{m}\!y_i^2-2\mu\sum_{i=1}^{m}\!y_i+m\mu^2\big\} \\
&\qquad -\frac{1}{2\sigma^2}
 \big\{\sum_{i=m+1}^{n}\!z_i^2-2\mu\sum_{i=m+1}^{n}\!z_i+(n-m)\mu^2\big\}.
\end{align*}

Because of the {iid} structure,
 the predictive distribution of the missing
data given ${\boldsymbol{\theta}}$ does not depend on the observed data.
Thus the $z_i$'s are observations from the truncated
normal distribution  
\begin{align} \label{EQ:tnorm}
&p({\mathbf{z}}|{\mathbf{y}},{\boldsymbol{\theta}}) 
=p({\mathbf{z}}|{\boldsymbol{\theta}})   \notag \\
&=\prod_{i=m+1}^{n} p({z_i}|{\boldsymbol{\theta}})  
 =\prod_{i=m+1}^{n}  \frac{\frac{1}{\sigma} \phi(\frac{z_i-\mu}{\sigma})}
       {1 - \Phi( \frac{R_i-\mu}{\sigma} )},
\qquad (z_i > R_i),
\end{align}
where $\phi(\cdot)$ and $\Phi(\cdot)$ are pdf and cdf of $N(0,1)$, respectively.
Using the following integral identities
\begin{align*}
\int\frac{z}{\sigma}\phi(\frac{z-\mu}{\sigma})dz 
     &= \mu \Phi(\frac{z-\mu}{\sigma})-\sigma\phi(\frac{z-\mu}{\sigma}),\\
\int\frac{z^2}{\sigma}\phi(\frac{z-\mu}{\sigma})dz 
    &= (\mu^2+\sigma^2)\Phi(\frac{z-\mu}{\sigma})
       -\sigma(\mu+z)\phi(\frac{z-\mu}{\sigma}),
\end{align*}
we have the expected log-likelihood
 at the $s$th step in the EM sequence: 
\begin{align*}
& Q({\boldsymbol{\theta}}|{\boldsymbol{\theta}}^{(s)}) \\
&= \int\log L^c({\boldsymbol{\theta}} | {\mathbf{y,z}}) 
            p({\mathbf{z}} | {\boldsymbol{\theta}}^{(s)}) d{\mathbf{z}} \\
&= C-\frac{n}{2}\log\sigma^2
 -\frac{1}{2\sigma^2} 
 \big\{\sum_{i=1}^{m}y_i^2-2\mu\sum_{i=1}^{m}y_i+m\mu^2\big\} \\
&\qquad -\frac{1}{2\sigma^2} \sum_{i=m+1}^{n} 
          \int_{R_i}^{\infty}(z_i^2-2\mu z_i+\mu^2)
                   p({z_i}|{\boldsymbol{\theta}}^{(s)})dz_i  \\
&= C-\frac{n}{2}\log\sigma^2 
    -\frac{1}{2\sigma^2}\{T_2 - 2\mu T_1 + m\mu^2\} \\
&\qquad -\frac{1}{2\sigma^2}\{ S_2^{(s)} - 2\mu S_1^{(s)} + (n-m)\mu^2 \},
\end{align*}
where $T_1$, $T_2$, $S_1^{(s)}$, and $S_2^{(s)}$ are given by
\begin{align*}
T_1 &= \sum_{i=1}^{m}y_i, \\
T_2 &= \sum_{i=1}^{m}y_i^2, \\
S_1^{(s)} &= \sum_{i=m+1}^{n} \int_{R_i}^{\infty} 
               z_i\,p({z_i}|{\boldsymbol{\theta}}^{(s)})dz_i  \\
          &= (n-m)\mu^{(s)} + \sum_{i=m+1}^{n} 
             \frac{\sigma^{(s)}\phi(\frac{R_i-\mu^{(s)}}{\sigma^{(s)}})}
                  {1-\Phi(\frac{R_i-\mu^{(s)}}{\sigma^{(s)}})},    \\
S_2^{(s)} &= \sum_{i=m+1}^{n} \int_{R_i}^{\infty} 
               z_i^2 p({z_i}|{\boldsymbol{\theta}}^{(s)})dz_i \\
          &= (n-m) \Big\{ (\mu^{(s)})^2 + {\sigma^2}^{(s)} \Big\} \\
          &\quad +\sum_{i=m+1}^{n} 
   \frac{(\mu^{(s)}+R_i)\sigma^{(s)}\phi(\frac{R_i-\mu^{(s)}}{\sigma^{(s)}})}
                  {1-\Phi(\frac{R_i-\mu^{(s)}}{\sigma^{(s)}})}.
\end{align*}

Differentiating 
the expected log-likelihood
$Q({\boldsymbol{\theta}}|{\boldsymbol{\theta}}^{(s)})$
with respect to $\mu$ and $\sigma^2$ and  solving for
$\mu$ and $\sigma^2$, we obtain the EM sequences
\begin{align} 
\mu^{(s+1)}   &= \frac{1}{n}\big\{T_1+S_1^{(s)}\big\}, \label{EQ:EMnormal1} \\
{\sigma^2}^{(s+1)}  &= \frac{1}{n}  \big\{T_2 + S_2^{(s)}\big\}
            -\frac{1}{n^2}\big\{T_1+S_1^{(s)}\big\}^2. \label{EQ:EMnormal2}
\end{align}

If we instead use the MCEM algorithm by
simulating $z_{m+1},\ldots,z_{n}$ from
the truncated normal
distribution $p({\mathbf{z}}|{\mathbf{y}},{\boldsymbol{\theta}}^{(s)})$
given by (\ref{EQ:tnorm}), then 
 the $Q({\boldsymbol{\theta}}|{\boldsymbol{\theta}}^{(s)})$ is replaced 
with the approximate expected log-likelihood
\begin{align*}
&\hat{Q}({\boldsymbol{\theta}}|{\boldsymbol{\theta}}^{(s)}) \\
&= \frac{1}{K}\sum_{k=1}^{K} 
   \log L^c ({\boldsymbol{\theta}}|{\mathbf{y}},{\mathbf{z}}^{(k)}) \\
&= C - \frac{n}{2}\log\sigma^2
     - \frac{1}{2\sigma^2}
   \Big\{T_2+\frac{1}{K}V_2^{(s)}
        -2\mu\big(T_1+\frac{1}{K}V_1^{(s)} \big)+n\mu^2\Big\} 
\end{align*}
where
\begin{align*}
&V_1^{(s)} = \sum_{k=1}^K\sum_{i=m+1}^n z_{i,k}, \\
&V_2^{(s)} = \sum_{k=1}^K\sum_{i=m+1}^n z_{i,k}^2, \\
&{\mathbf{z}}^{(k)}=(z_{m+1,k},\ldots,z_{n,k}), 
\end{align*}
and $z_{i,k}$ is from
%----------------
\begin{equation*} 
 p({z_{i,k}}|{\boldsymbol{\theta}}^{(s)})
= \frac{\frac{1}{\sigma} \phi(\frac{z_{i,k}-\mu}{\sigma})}
       {1 - \Phi( \frac{R_i-\mu}{\sigma} )}
\qquad (z_{i,k} > R_i),
\end{equation*}
%----------------
for $i=m+1,\ldots,n$.
We then obtain the MCEM sequences by differentiating the
$\hat{Q}({\boldsymbol{\theta}}|{\boldsymbol{\theta}}^{(s)})$
\begin{align} 
\mu^{(s+1)}  & = \frac{1}{n}\big\{T_1 + \frac{1}{K}V_1^{(s)}\big\},
              \label{EQ:MCEMnormal1} \\
{\sigma^2}^{(s+1)} &= \frac{1}{n}\big\{T_2 + \frac{1}{K}V_2^{(s)}\big\}
                    - \frac{1}{n^2}\big\{T_1 + \frac{1}{K}V_1^{(s)}\big\}^2.
              \label{EQ:MCEMnormal2} 
\end{align}
%================
This is merely an example of the MCEM algorithm since 
the ordinary EM algorithm applies.

%-------------------------------------
\subsection{The Laplace  distribution}
%-------------------------------------
Let $Y_1,\ldots,Y_m$ and $Z_{m+1},\ldots,Z_n$ be
iid Laplace random variables with 
$\boldsymbol{\theta}=(\mu,\sigma)$,
where the pdf is
$$
f(x|\boldsymbol{\theta})
  = \frac{1}{2\sigma} \exp\Big(-\frac{|x-\mu|}{\sigma}\Big),
\qquad  \sigma>0.
$$
Then the complete-data log-likelihood is
\begin{align*}
&\log L^c({\boldsymbol{\theta}} | {\mathbf{y,z}})  \\
&= C - n\log\sigma
- \frac{1}{\sigma} \sum_{i=1}^{m} |y_i-\mu|
- \frac{1}{\sigma} \sum_{i=m+1}^{n} |z_i-\mu|.
\end{align*}
Because of the {iid} structure, 
the predictive distribution of the missing data given ${\boldsymbol{\theta}}$ 
does not depend on the observed data.
Thus the $z_i$'s are observations from the truncated
Laplace distribution  
\begin{align*}
& p({\mathbf{z}}|{\mathbf{y}},{\boldsymbol{\theta}}) 
 =p({\mathbf{z}}|{\boldsymbol{\theta}})     \\
&=\prod_{i=m+1}^{n} p({z_i}|{\boldsymbol{\theta}})
 =\prod_{i=m+1}^{n}
 \frac{f(z_i|{\boldsymbol{\theta}})}{1-F(R_i|{\boldsymbol{\theta}})},
\qquad (z_i > R_i),
\end{align*}
where $F(\cdot)$ is the cdf of Laplace random variable.
Then at the $s$th step in the EM sequence, we have
the expected log-likelihood
\begin{align*}
&Q({\boldsymbol{\theta}}|{\boldsymbol{\theta}}^{(s)}) \\
&= \int\log L^c({\boldsymbol{\theta}} | {\mathbf{y,z}}) 
            p({\mathbf{z}} | {\boldsymbol{\theta}}^{(s)}) d{\mathbf{z}} \\
&= C - n\log\sigma
- \frac{1}{\sigma} \sum_{i=1}^{m}  |y_i-\mu| 
- \frac{1}{\sigma} \sum_{i=m+1}^{n}\int_{R_i}^{\infty} |z_i-\mu| \,
            p({z_i}|{\boldsymbol{\theta}}^{(s)}) d{z_i}.
\end{align*}
The computation of the above integration part is very complex.
We can overcome this difficulty by using MCEM approach.
The approximate expected log-likelihood is 
\begin{align*}
&\hat{Q}({\boldsymbol{\theta}}|{\boldsymbol{\theta}}^{(s)}) \\
&=\frac{1}{K} \sum_{k=1}^{K} \log L^c({\boldsymbol{\theta}} | 
                              {\mathbf{y}},{\mathbf{z}}^{(k)}) \\
&=C - n\log\sigma - \frac{1}{\sigma} \frac{1}{K}\sum_{k=1}^{K}
    \Big\{ \sum_{i=1}^{m} |y_i-\mu| + \sum_{i=m+1}^{n}|z_{i,k}-\mu| \Big\},
\end{align*}
where ${\mathbf{z}}^{(k)}=(z_{m+1,k},\ldots,z_{n,k})$ is from
$p({\mathbf{z}}|{\boldsymbol{\theta}}^{(s)})$. 
Note that it is easy to simulate a truncated Laplace random variable by
using the inverse transformation method; see Appendix.
Using this, we obtain the MCEM sequences:
\begin{align}  
\mu^{(s+1)}
&= {\mathrm{median}}
  ({\mathbf{y}}\otimes K,~
   {\mathbf{z}}^{(1)},\ldots,{\mathbf{z}}^{(K)})
                                     \label{EQ:MCEMLaplace1} \\
\sigma^{(s+1)}
&= \frac{1}{n} 
   \Big\{ \sum_{i=1}^{m} |y_i-\mu^{(s+1)}| + 
          \frac{1}{K}\sum_{k=1}^{K}\sum_{i=m+1}^{n}|z_{i,k}-\mu^{(s+1)}|
   \Big\},     \label{EQ:MCEMLaplace2}
\end{align}
where ${\mathbf{y}}\otimes K = (\mathbf{y},\ldots,\mathbf{y})$, that is, 
$K$ replications of $\mathbf{y}$.

%-------------------------------------
\subsection{The Rayleigh distribution}
%-------------------------------------
Let $Y_1,\ldots,Y_m$ and $Z_{m+1},\ldots,Z_n$ be
iid Rayleigh random variables with the pdf:
$$
f(x|\beta)
  = \frac{x}{\beta^2} \exp\big(-\frac{x^2}{2\beta^2}\big),
\quad x>0,~ \beta>0.
$$
Then the complete-data log-likelihood is
\begin{align*}
\log L^c({{\beta}} | {\mathbf{y,z}}) 
&= C-2n\log\beta 
     +\sum_{i=1}^{m}\log y_i- \frac{1}{2\beta^2}\sum_{i=1}^{m}y_i^2 \\
&\qquad+\sum_{i=m+1}^{n}\log z_i - \frac{1}{2\beta^2}\sum_{i=m+1}^{n}z_i^2.
\end{align*}
The predictive distribution of the missing data given ${{\beta}}$ 
does not depend on the observed data.
Thus the $z_i$'s are observations from the truncated Rayleigh distribution
\begin{align*} 
& p({\mathbf{z}}|{\mathbf{y}},{{\beta}})  \\
& =\prod_{i=m+1}^{n} p({z_i}|{{\beta}})
  =\prod_{i=m+1}^{n}\frac{1}{\beta^2}
                  z_i\exp\big(\frac{R_i^2-z_i^2}{2\beta^2}\big),
\qquad (z_i > R_i).
\end{align*}
Then at the $s$th step in the EM sequence, we have
the expected log-likelihood
\begin{align*} 
& Q(\beta|\beta^{(s)})  \\
&= \int\log L^c(\beta | {\mathbf{y,z}}) 
            p({\mathbf{z}} | \beta^{(s)}) d{\mathbf{z}} \\
&= C - 2n\log\beta
  +\sum_{i=1}^{m}  \log y_i - \frac{1}{2\beta^2}\sum_{i=1}^{m}  y_i^2 \\
&\qquad +\sum_{i=m+1}^{n} \int_{R_i}^{\infty} 
  (\log z_i-\frac{1}{2\beta^2}z_i^2) p({z_i} | \beta^{(s)}) d{z_i}.
\end{align*}
The calculation of the above integration part
does not have a closed form.
Using MCEM, we have
the approximate expected log-likelihood
\begin{align*}
& \hat{Q}({{\beta}}|{{\beta}}^{(s)}) \\
&= \frac{1}{K}\sum_{k=1}^{K}
   \log L^c({{\beta}}|{\mathbf{y}},{\mathbf{z}}^{(k)})\\
&= C - 2n\log\beta
+\sum_{i=1}^{m}  \log y_i - \frac{1}{2\beta^2}\sum_{i=1}^{m}  y_i^2 \\
&\qquad+\frac{1}{K}\sum_{k=1}^{K}\sum_{i=m+1}^{n}\log z_{i,k} 
-\frac{1}{2\beta^2}\frac{1}{K}\sum_{k=1}^{K} \sum_{i=m+1}^{n}z_{i,k}^2,
\end{align*}
where ${\mathbf{z}}^{(k)}=(z_{m+1,k},\ldots,z_{n,k})$ is from
$p({\mathbf{z}}|{\beta}^{(s)})$.
Using the inverse transformation method,
we can simulate a truncated Rayleigh random variable $X$ 
$$
X = \sqrt{R^2 - {2\beta^2}\log U},
$$
where $U$ is a ${\mathrm{uniform}}\,(0,1)$ random variable.
We then obtain the MCEM sequences by differentiating 
$\hat{Q}({{\beta}}|{{\beta}}^{(s)})$
\begin{equation}  \label{EQ:MCEMRayleigh}
\beta^{(s+1)} 
=\sqrt{\frac{1}{2n} \Big\{T_2 + \frac{1}{K}V_2^{(s)}\Big\} },
\end{equation}
where 
\begin{align*}
&T_2       = \sum_{i=1}^{m}y_i^2, \\ 
&V_2^{(s)} = \sum_{k=1}^K\sum_{i=m+1}^n z_{i,k}^2.
\end{align*}
%================

%============================
\section{{Illustrative Examples}\label{SEC:Examples}}
%============================
This section provides three numerical examples of parameter estimation 
for the normal, Laplace and Rayleigh distributions
 using the EM algorithms.

%-------------------------------------
\subsection*{Example 1: censored normal sample}
%-------------------------------------
Let us consider the data presented earlier by
Gupta~\cite{Gupta:1952} in which, out of $N=10$,
the largest three have been censored. 
The Type-II right censored sample is as follows:
$$
1.613,~ 1.644,~ 1.663,~ 1.732,~ 1.740,~ 1.763,~ 1.778.
$$

In this case, Gupta~\cite{Gupta:1952} computed the
estimates of the mean and the standard deviation
by three different methods, viz.\
(i) best linear ($\hat{\mu}=1.746,\hat{\sigma}=0.101$),
(ii) alternative linear  ($\hat{\mu}=1.748,\hat{\sigma}=0.094$), 
and
(iii) maximum likelihood  ($\hat{\mu}=1.742,\hat{\sigma}=0.072$).
His calculation of the MLE seems to be incorrect.
The new calculation of the MLE is 
$\hat{\mu}=1.742$ and $\hat{\sigma}=0.079$.

We use the EM sequences from (\ref{EQ:EMnormal1}) and (\ref{EQ:EMnormal2}).
Table~\ref{EM:norm1} presents the iteration sequence of the implementation
of the EM algorithm for this problem.
Starting values are chosen by (i) taking the sample mean 
and sample variance of the uncensored data
($\mu^{(0)}=1.7$, ${\sigma^2}^{(0)}=0.004$)
 and (ii) selecting arbitrary numbers
(for example, $\mu^{(0)}=0$, ${\sigma^2}^{(0)}=1$).
We obtain the same result in both cases up to the third decimal point
after about 10 iterations. 
 
Next, we use the MCEM  sequences
from (\ref{EQ:MCEMnormal1}) and (\ref{EQ:MCEMnormal2}).
Table~\ref{MCEM:norm1} presents the iteration sequence of the implementation
of the MCEM algorithm.
The algorithm was run with $K=50,000$ for 15 iterations
with different starting values, yielding
the same results as the MLE up to the third decimal place.

\bigskip
\fbox{Table~\ref{EM:norm1} around here}

\bigskip
\fbox{Table~\ref{MCEM:norm1} around here}

 \clearpage
%-------------------------------------
\subsection*{Example 2: censored Laplace sample}
%-------------------------------------
Let us consider the data presented earlier by
Balakrishnan~{\em et al.}~\cite{Balakrishnan:1996}
(simulated with $\mu=50$ and $\sigma=5$) in which,
out of $N=20$ observations, the largest two have been censored.
The Type-II right-censored sample thus obtained is as follows:
\begin{center}
\begin{tabular}{r@{~~}r@{~~}r@{~~}r@{~~}r}
 32.00692, &  37.75687,& 43.84736,& 46.26761,& 46.90651, \\
 47.26220, &  47.28952,& 47.59391,& 48.06508,& 49.25429, \\
 50.27790, &  50.48675,& 50.66167,& 53.33585,& 53.49258, \\
 53.56681, &  53.98112,& 54.94154.&          
\end{tabular}
\end{center}

In this case, Balakrishnan~{\em et al.}~\cite{Balakrishnan:1996}
computed the BLUEs of $\mu$ and $\sigma$ as 
$\hat{\mu}=49.56095$ and $\hat{\sigma}=4.81270$.
The MLE is $\hat{\mu}=49.76609$ and $\hat{\sigma}=4.68761$.

We use the MCEM  sequences
from (\ref{EQ:MCEMLaplace1}) and (\ref{EQ:MCEMLaplace2}).
Table~\ref{MCEM:Laplace} presents the iteration sequence of the implementation
of the MCEM algorithm.
The algorithm was run with $K=50,000$ for 5 iterations
with the starting value ($\mu^{(0)}=0,\sigma^{(0)}=1$), yielding
the same results as the MLE up to third decimal place. 
When compared to the BLUE, our result is closer to the MLE.

\bigskip
\fbox{Table~\ref{MCEM:Laplace} around here}

%-------------------------------------
\subsection*{Example 3: censored Rayleigh sample}
%-------------------------------------
We simulated a data set with $\beta=5$ in which,
out of $N=20$ observations, the largest five have been
censored. 
The Type-II right censored sample thus obtained is as follows:
\begin{center}
\begin{tabular}{r@{~~}r@{~~}r@{~~}r@{~~}r}
1.950, & 2.295, & 4.282, & 4.339, & 4.411, \\
4.460, & 4.699, & 5.319, & 5.440, & 5.777, \\
7.485, & 7.620, & 8.181, & 8.443, &10.627.
\end{tabular}
\end{center}

We use the MCEM  sequences
from (\ref{EQ:MCEMRayleigh}).
Table~\ref{MCEM:Rayleigh} presents the iteration sequence of 
the implementation of the MCEM algorithm.
Two different starting values
($\beta^{(0)}=1$ and $\beta^{(0)}=100$)
are chosen to show that the MCEM
is very insensitive to the choice of starting value.
This iteration sequence shows that the MCEM converges vary fast.
The algorithm was run with $K=50,000$ for 10 iterations.
We obtain $\hat{\beta}=6.1324$ and $\hat{\beta}=6.1332$
 with different starting values.
The MLE is $\hat{\beta}=6.1341$.
The results are the same as the MLE up to the second decimal place.

\bigskip
\fbox{Table~\ref{MCEM:Rayleigh} around here}

\clearpage
%===========================
\section*{Appendix}
%===========================
%-------------------------------------
\subsection*{Simulation of truncated normal random variable}
%-------------------------------------
Let $U$ be a uniform $(0,1)$ random variable.
For any continuous cdf $F(\cdot)$ if we define the random variable
$X$ by $X=F^{-1}(U)$, then the random variable
has distribution function $F(\cdot)$; see~\cite{Ross:2000}.
This method is called the {\em inverse transformation method}.

Using this method, we have the following 
truncated normal random variable $X$:
$$
X=\mu+\sigma\,\Phi^{-1}\Big\{ \big((1-\Phi(r)\big)U + \Phi(r)\Big\},
$$
where $r=(R-\mu)/\sigma$ and $U \sim U(0,1)$.

%-------------------------------------
\subsection*{Simulation of truncated Laplace random variable}
%-------------------------------------
Using the inverse transformation method, we have the following
truncated Laplace random variable $X$:
%-------------------------------
\itemindent 0em \leftmargini 2em
\begin{enumerate}
\item[(i)] $R \ge \mu$ \\
$X = R - \sigma\log U$, 
\item[(ii)] $R  < \mu$ \\
$
X=\left\{ 
\begin{array}{l@{\quad}l}
\mu + \sigma\log \big\{2U+(1-U)\exp(r) \big\} 
                    & {\mathrm{if~}} U \le H \\
\mu - \sigma\log \big\{2(1-U)-(1-U)\exp(r) \big\} 
                    & {\mathrm{if~}} U > H,
\end{array}
\right.
$
\end{enumerate}
where $r=(R-\mu)/\sigma$ and $H=(1-e^{r})/(2-e^{r})$.

\clearpage
%%==================================================================
\nocite{Robert/Casella:1999}
%% \nocite{R:2005}
%% \bibliographystyle{plain}
%% \bibliography{/home/cspark/MyFiles/library/TeX/bib/STAT,%
%%  /home/cspark/MyFiles/library/TeX/bib/ENG}

%%==================================================================

\clearpage
\begin{table}
\begin{center}
\caption{Iterations of EM  with normal sample.
         \label{EM:norm1}}
\medskip
\begin{tabular}{r@{\quad}llc@{\quad}ll}
\hline
$s$ & $~~\mu^{(s)}$ & $~~{\sigma}^{(s)}$
   && $~~\mu^{(s)}$ & $~~{\sigma}^{(s)}$ \\
\hline
0 &    1.7    & $0.004^{1/2}$ &&  0       & 1      \\
1 &    1.7358 &  0.0702       &&  1.8467  & 0.2968 \\
2 &    1.7397 &  0.0754       &&  1.8058  & 0.1931 \\
3 &    1.7411 &  0.0775       &&  1.7761  & 0.1370 \\
4 &    1.7418 &  0.0784       &&  1.7593  & 0.1070 \\
5 &    1.7420 &  0.0788       &&  1.7504  & 0.0919 \\
6 &    1.7421 &  0.0790       &&  1.7459  & 0.0848 \\
7 &    1.7422 &  0.0791       &&  1.7439  & 0.0816 \\
8 &    1.7422 &  0.0791       &&  1.7429  & 0.0802 \\
9 &    1.7422 &  0.0791       &&  1.7425  & 0.0796 \\
10&    1.7422 &  0.0791       &&  1.7424  & 0.0793 \\
11&    1.7422 &  0.0791       &&  1.7423  & 0.0792 \\
12&    1.7422 &  0.0791       &&  1.7423  & 0.0792 \\
\hline
\end{tabular}
\end{center}
\end{table}

\clearpage
\begin{table}[h]
\begin{center}
\caption{Iterations of MCEM ($K=50,000$) with normal sample.
 \label{MCEM:norm1}}
\medskip
\begin{tabular}{r@{\quad}llc@{\quad}ll}
\hline
$s$ & $~~\mu^{(s)}$ & $~~{\sigma}^{(s)}$
   && $~~\mu^{(s)}$ & $~~{\sigma}^{(s)}$ \\
\hline
0 &    1.7    & $0.004^{1/2}$&& 0      & 1      \\
1 &    1.7363 &  0.0708      && 1.8472 & 0.2976  \\
2 &    1.7398 &  0.0756      && 1.8061 & 0.1938  \\
3 &    1.7412 &  0.0777      && 1.7763 & 0.1375  \\
4 &    1.7417 &  0.0784      && 1.7593 & 0.1070  \\
5 &    1.7420 &  0.0788      && 1.7503 & 0.0918  \\
6 &    1.7421 &  0.0789      && 1.7459 & 0.0847  \\
7 &    1.7422 &  0.0791      && 1.7439 & 0.0816  \\
8 &    1.7422 &  0.0791      && 1.7429 & 0.0802  \\
9 &    1.7423 &  0.0792      && 1.7426 & 0.0796  \\
10&    1.7423 &  0.0792      && 1.7424 & 0.0794  \\
11&    1.7423 &  0.0792      && 1.7423 & 0.0793  \\
12&    1.7423 &  0.0792      && 1.7423 & 0.0793  \\
13&    1.7422 &  0.0792      && 1.7422 & 0.0792  \\
14&    1.7422 &  0.0791      && 1.7422 & 0.0791  \\
15&    1.7423 &  0.0792      && 1.7423 & 0.0792  \\
\hline
\end{tabular}
\end{center}
\end{table}

\clearpage
\begin{table}[htb]
\begin{center}
\caption{Iterations of MCEM ($K=50,000$) with Laplace sample.
\label{MCEM:Laplace}}
\medskip
\begin{tabular}{r@{\quad}ll}
\hline
$s$ & $~~\mu^{(s)}$ & $~~{\sigma}^{(s)}$ \\
\hline
0 & 0       & 1      \\
1 & 49.7661 & 4.3189      \\
2 & 49.7661 & 4.6493      \\
3 & 49.7661 & 4.6844      \\
4 & 49.7661 & 4.6884      \\
5 & 49.7661 & 4.6882      \\
\hline
\end{tabular}
\end{center}
\end{table}

\clearpage
\begin{table}[htb]
\begin{center}
\caption{Iterations of MCEM ($K=50,000$) with Rayleigh sample.
         \label{MCEM:Rayleigh}}
\medskip
\begin{tabular}{r@{\quad}ll}
\hline
$s$ & $~~{\beta}^{(s)}$ & $~~{\beta}^{(s)}$ \\
\hline
0 & 1      & 100         \\
1 & 5.3358 & 50.2854      \\
2 & 5.9450 & 25.7063      \\
3 & 6.0900 & 13.9290      \\
4 & 6.1254 &  8.7677      \\
5 & 6.1321 &  6.8879      \\
6 & 6.1318 &  6.3287      \\
7 & 6.1329 &  6.1827      \\
8 & 6.1353 &  6.1478      \\
9 & 6.1326 &  6.1358      \\
10& 6.1324 &  6.1332      \\
\hline
\end{tabular}
\end{center}
\end{table}

%%===============================================
\end{document}